\documentclass[aps,pra,twocolumn,superscriptaddress]{revtex4}
\usepackage{amsfonts,epsfig}
\usepackage{amsmath}
\usepackage{amssymb}
\usepackage{amsfonts}
\usepackage{amsthm}
\usepackage{mathrsfs}
\usepackage{color,verbatim,graphics}
\usepackage{url}
\usepackage{multirow}

\newcommand{\refereqn}[1]{(\ref{#1})}

\newcommand{\ket}[1]{\left |  #1 \right \rangle}
\newcommand{\bra}[1]{\left \langle #1  \right |}

\def\proj#1{\ket{#1}\!\bra{#1}}

\newcommand{\tr}[2][ ]{\text{Tr}_{#1}\!\left( #2 \right)}

\def\id{{\mathbb I}}

\newcommand{\be}{\begin{equation}}
\newcommand{\ee}{\end{equation}}
\newcommand{\ba}{\begin{eqnarray}}
\newcommand{\ea}{\end{eqnarray}}
\newcommand{\ban}{\begin{eqnarray*}}
\newcommand{\ean}{\end{eqnarray*}}

\newtheorem*{theorem*}{Theorem}

\newcommand{\bracket}[3]{\langle#1|#2|#3\rangle}

\newcommand{\ketbra}[2]{|#1\rangle\hspace{-2.4pt}\langle #2| }

\begin{document}

\title{Robust and versatile black-box certification of quantum devices}

\author{Tzyh Haur Yang}
\affiliation{Centre for Quantum Technologies, National University of Singapore, 3 Science drive 2, Singapore 117543}
\author{Tam\'as V\'ertesi}
\affiliation{Institute for Nuclear Research, Hungarian Academy of Sciences, H-4001 Debrecen, P.O. Box 51, Hungary}
\author{Jean-Daniel Bancal}
\affiliation{Centre for Quantum Technologies, National University of Singapore, 3 Science drive 2, Singapore 117543}
\author{Valerio Scarani}
\affiliation{Centre for Quantum Technologies, National University of Singapore, 3 Science drive 2, Singapore 117543}
\affiliation{Department of Physics, National University of Singapore, 2 Science drive 3, Singapore 117542}
\author{Miguel Navascu\'es}
\affiliation{School of Physics, University of Bristol, Tyndall Avenue, Bristol BS8 1TL (U.K.)}

\begin{abstract}
Self-testing refers to the fact that, in some quantum devices, both states and measurements can be assessed in a black-box scenario, on the sole basis of the observed statistics, i.e. without reference to any prior device calibration. Only a few examples of self-testing are known, and they just provide non-trivial assessment for devices performing unrealistically close to the ideal case. We overcome these difficulties by approaching self-testing with the semi-definite programming hierarchy for the characterization of quantum correlations. This allows us to improve dramatically the robustness of previous self-testing schemes -e.g.: we show that a CHSH violation larger than 2.57 certifies a singlet fidelity of more than $70\%$. In addition, the versatility of the tool brings about self-testing of hitherto impossible cases, such as robust self-testing of non-maximally entangled two-qutrit states in the CGLMP scenario.
\end{abstract}

\maketitle

%%%%%%%%%%%%%%%%%%%%%%%%%%%%%%%%%%%%%%%%%%%%%%
\textit{Introduction} - The validation and certification of sources and measurement apparatuses constitutes a fundamental step of science and technology. One does not buy the elements to set up an experiment without first assessing their quality; and one should not make claims about the final results of an experiment without several checks. Usually, a variety of assumptions go into these procedures. For instance, the certification of a device often depends on the fact that other devices are properly calibrated~\cite{lundeen09}. In the last few years, it has been noticed that tasks like quantum key distribution \cite{qkd1} and random number generation \cite{random1,random2} can be validated based only on minimal assumptions and on the statistics observed a posteriori. The idea consists in looking for statistics that violate Bell inequalities; the minimal assumptions that go into this so-called \textit{device-independent assessment} are essentially no-signaling (which could in principle be guaranteed by putting a sufficient distance between the devices) and measurement independence (i.e. the possibility of performing different measurements on the same setup, a cornerstone of the scientific method) \cite{slovakia,review}.

Rather than certifying that some device can accomplish a task, one may want to \textit{certify the device itself}, which in turn would provide certification for any possible further task one may want to perform with it. For instance, if the device is a source, this would amount to performing a ``blind tomography'' where measurement devices are treated as black boxes. It has long been known that this is possible in some specific and ideal cases. Famously, if the CHSH inequality~\cite{CHSH} is violated at its maximal value $2\sqrt{2}$, the devices are certified to be performing complementary measurements on two effective qubits in the maximally entangled state~\cite{sw87,pr92,Tsirelson93}. Another criterion that certifies the same state and measurements was put forward by Mayers and Yao, who called the whole task \textit{self-testing of quantum apparatuses}~\cite{yao}.

In addition to being tailored for two-qubit singlet, these pioneering works are unapplicable to real-world devices because they only discuss the statistics of the ideal case. A first step towards the resolution of this issue was taken when several self-testing schemes were shown to be ``robust" (or ``rigid'')~\cite{endnotewerner,magniez,mckague,miller}; the most advanced of these results applies to a multiple-copy scenario and certifies the state as a resource for universal quantum computation~\cite{reichardt}. Despite the name, however, these results tolerate only tiny deviations from the ideal case. Take again the certification of the two-qubit singlet based on the CHSH inequality: even for the largest reported experimental violation, which is $2.827 \pm 0.0017$~\cite{Christensen13} i.e. only $0.1\%$ away from the ideal value, none of the ``robust'' self-testing approaches quoted above provide a nontrivial bound on the singlet fidelity.

One may surmise that this could be an intrinsic limitation on the ambitious task of self-testing. Here, we show that this is not the case: we demonstrate that a CHSH violation of $2.827$ is only compatible with a singlet fidelity larger than $99.83\%$. This real-life robustness is only one of the benefits of the method that we introduce. Indeed, our approach formalizes the idea of \textit{swapping black boxes with trusted systems} \cite{yao} with the semidefinite characterization of quantum correlations~\cite{navascues}, which makes it especially versatile. We demonstrate this explicitely with several examples, all of which are robust. Notably, we describe self-testing of qutrit states with ternary outcome measurements, which would not be possible with previous techniques.

In most self-testing works, the assumption is made that the tested devices behave independently and in an identical way (i.i.d.) over the runs. This assumption may sound problematic, as it may fail in real situations (e.g. if a source is drifting). Fortunately, tools have been developed to deal with the general case of Bell-based tests where each realization of the box can be different from the previous one and may even depend on all previous operations effected on the system ~\cite{bchkp,gill,Zhang11}. With these tools, the results obtained with i.i.d. hold true in the general case, in the asymptotic limit of infinitely many runs. In this paper, we work only in that limit, so we take i.i.d. for granted in the rest of the paper.

For clarity of presentation, we now introduce our method with the basic example of two-qubit singlet state certification via the CHSH inequality. A few other applications are discussed in the remainder of the paper, and many more are left for future work.

\begin{figure}
\includegraphics[width=0.4\textwidth]{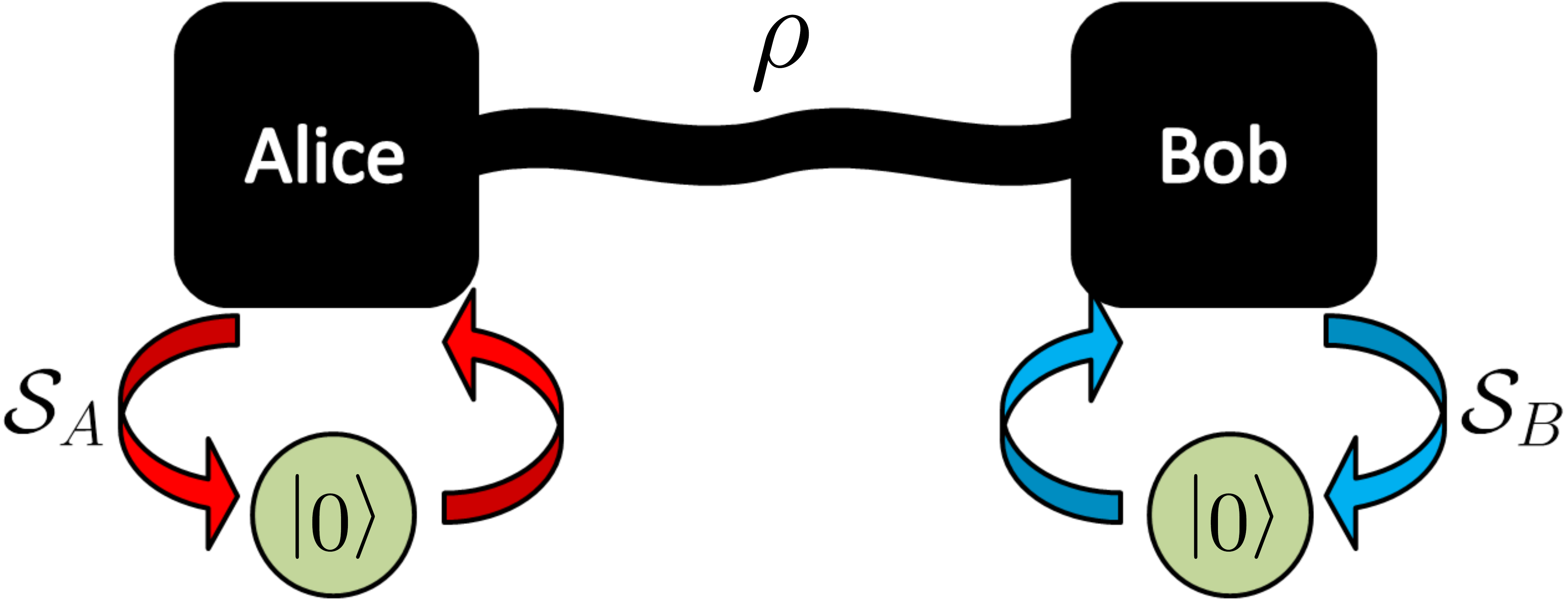}
\caption{The swap concept: Characteristics of black boxes are assessed by considering the effect of swap operations between these black boxes and trusted systems (initialized in the state $\ket{0}$ here).}
\label{fig:fullSwap}
\end{figure}

\textit{Bound on the singlet fidelity from CHSH} - Let us consider a bipartite experiment with binary inputs $x,y\in\{0,1\}$ and binary outputs $a,b\in\{0,1\}$. After querying the boxes a large number of times, one can reconstruct the measurement statistics $P(ab|xy)$; the CHSH inequality is violated if $\mathcal{B}_\text{CHSH} = \sum_{abxy} (-1)^{a+b+xy}P(ab|xy)>2$ \cite{endnotedet}. If a violation is observed, the measured state must be entangled, and it must even be a maximally-entangled singlet state $\ket{\psi^-}=(\ket{01}-\ket{10})/\sqrt{2}$ if the violation is maximal. Our goal is to quantify how far from the singlet the state can be, in terms of fidelity, when the violation is not maximal.
Since nothing guarantees that the state in the boxes is a two-qubit state, one must clarify what the fidelity with the singlet means at all. The idea of self-testing consists in swapping part of the content of the black boxes into a trusted system (in this case two qubits) initially prepared in a suitable dummy state. The singlet fidelity of the final two-qubit state is then well-defined.

Specifically, let the trusted auxiliary qubits $A'$ and $B'$ be prepared in the state $\ket{0}$. Then some local unitaries ${\cal S}_{AA'}$ and ${\cal S}_{BB'}$ are applied between these trusted systems and their respective boxes, as shown in figure~\ref{fig:fullSwap}. Such hypothetical operations leave the trusted systems in the state
\begin{equation}\label{eq:swappedstate}
\rho_\text{swap}=\text{tr}_{AB}\left[{\cal S} \rho_{AB}\otimes\ketbra{00}{00}_{A'B'} {\cal S}^\dag \right],
\end{equation}
where ${\cal S}={\cal S}_{AA'}\otimes {\cal S}_{BB'}$. This operation is a local isometry from the black box to the trusted space, as usually considered in self-testing. One wants to choose ${\cal S}$ such that $F=\bra{\psi^-}\rho_\text{swap}\ket{\psi^-}$ is large, possibly maximal.

It is crucial to stress that this isometry is the virtual procedure that allows one to define a figure of merit, \textit{not} a procedure that must be implemented in the lab for the certification to be possible. All that needs to be done in the lab is to collect the data that lead to reconstructing $P(ab|xy)$. Therefore, the alleged swap operation ${\cal S}$ itself must be defined, and its performance evaluated, from the observed statistics and the belief that whatever happens can be described within the framework of quantum theory. The latter tells us that, to any input $x$ of Alice, there correspond in the box one hermitian operators $\Pi^x_{a}$ for each outcome $a$, which can be taken as a projector since the dimension of the system being measured is not restricted.
The same holds for Bob. Based on these existing projectors, it is convenient to define the hermitian and unitary operators $A_x = \Pi^x_{0}-\Pi^x_{1}$ and $B_y = \Pi^y_{0}-\Pi^y_{1}$. Also, we describe the ideal state as \begin{equation}\label{eq:q2}
\ket{\overline\psi} = \cos\left(\frac{\pi}{8}\right)\ket{\phi^+} + \sin\left(\frac{\pi}{8}\right)\ket{\psi^+}\,,
\end{equation} which is maximally entangled and therefore equivalent to $\ket{\psi^-}$ up to local unitaries. This is chosen for convenience of notation since this states achieves $\mathcal{B}_\text{CHSH}=2\sqrt{2}$ for the operators
\begin{equation}\label{eq:q1}
\overline{A_0} = \overline{B_0} = \sigma_z,\ \overline{A_1} = \overline{B_1} = \sigma_x\,.
\end{equation}

All the framework is set. In order to guess a good construction for ${\cal S}$, we get inspiration from the ideal case. If the system in each box were indeed a qubit, the swap operations could be realized by combining three CNOT gates~\cite{nielsen_chuang}. Further, using \eqref{eq:q1}, the CNOT that has $A$ as target and $A'$ as control can be written as $\overline{U}_{AA'}=\openone\otimes\ketbra{0}{0} + \overline{A_1}\otimes\ketbra{1}{1}$; the CNOT with reversed roles can be written as $\overline{V}_{AA'} = \frac{\openone+\overline{A_0}}2\otimes\openone + \frac{\openone-\overline{A_0}}2\otimes\sigma_x$. Having noticed this, for the untrusted case we can tentatively define
\ba {\cal S}_{AA'} = U_{AA'} V_{AA'} U_{AA'} \label{eq:swapgen}\ea with
\begin{equation}
\begin{split}\label{eq:UV}
U_{AA'} &= \openone\otimes\ketbra{0}{0} + A_1\otimes\ketbra{1}{1}\\
V_{AA'} &= \frac{\openone+A_0}2\otimes\openone + \frac{\openone-A_0}2\otimes\sigma_x,
\end{split}
\end{equation}
and similarly for Bob. These operations are unitary for all $A_0$ and $A_1$ unitary and hermitian. Obviously, their actual action may differ from perfect swaps. For instance, suppose that the states and measurements in the boxes are equivalent to \eqref{eq:q2} and \eqref{eq:q1} up to local unitaries: the swapped state is always found to be $\rho_\text{swap} = \ketbra{\overline \psi}{\overline \psi}$ rather than its unitary equivalent. In other words, on maximally entangled two-qubit states and complementary measurements, this ${\cal S}$ act as ``clever swap" that compensates for local unitaries to produce always the desired output state.

\begin{figure}
\includegraphics[width=0.5\textwidth]{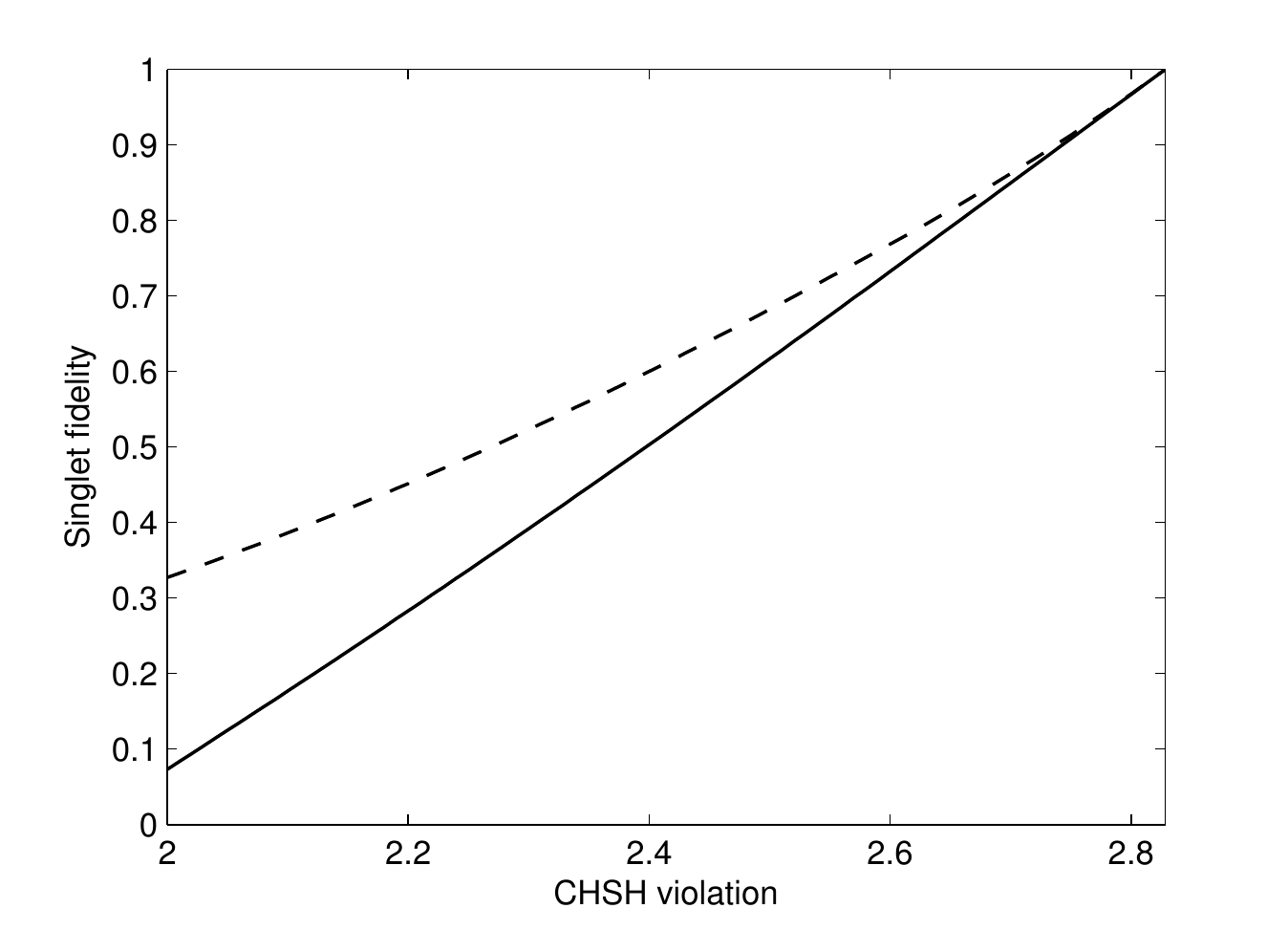}
\caption{Minimal singlet fidelity as a function of CHSH violation. The solid line denotes a lower bound on the fidelity for generic boxes; the dashed one a lower bound for isotropic boxes. Improved bounds are presented in~\cite{fulllength} using optimized swap operators.}
\label{fig:CHSH}
\end{figure}

Now that ${\cal S}$ is given explicitly in terms of $A_0$, $A_1$, $B_0$ and $B_1$, the partial trace \eqref{eq:swappedstate} can be formally computed \cite{endnoteuv}: the entries of $\rho_\text{swap}$ are given by linear combinations of correlation terms from the set $c=\{c_{\openone}=\text{tr}(\rho_{AB}\openone),c_{A_0}=\text{tr}(\rho_{AB}A_0),\dots, c_{A_0A_1B_0}=\text{tr}(\rho_{AB} A_0A_1B_0),\ldots\}$. The fidelity $\overline{F}=\bra{\overline\psi}\rho_\text{swap}\ket{\overline\psi}$ is thence a linear combination of these moments, and so is the CHSH expression. This allows one to relate the observed CHSH violation to the overlap. Since any such moments that proceed from a quantum realization satisfy some semidefinite constraints~\cite{navascues,vandenberghe}, a lower bound on the fidelity of the swapped state is obtained by solving the following semi-definite program (SDP):
\begin{align}
	f=\min \;\; &\bracket{\overline{\psi}}{\rho_\text{swap}}{\overline{\psi}} \nonumber\\
	\textrm{such that }&\hspace{0.2cm} c\in {\cal Q}_n \nonumber\\
			    & c_{A_0B_0}+c_{A_1B_0}+c_{A_0B_1}-c_{A_1B_1}=\mathcal{B}_\text{CHSH},
\end{align}
where ${\cal Q}_n$ is a relaxation of the quantum set. We run the SDP for various values of $\mathcal{B}_\text{CHSH}$. The result is the lowest curve of figure~\ref{fig:CHSH}. It is now simple to add constraints: for instance, the actual statistics may correspond to isotropic boxes, i.e. $c_{A_0B_0}=c_{A_1B_0}=c_{A_0B_1}=-c_{A_1B_1}$ and $c_{A_x}=c_{B_y}=0$, and these conditions can be added to the SDP.

\textit{Remarks on the method} - The crucial element of our method is the swap operator $\cal S$. Once expressed from the expected behavior of the boxes, and guaranteed to be unitary, the fidelity becomes a linear combination of moments $c$, which allow its optimization by SDP.
The observed statistics enter this SDP as constraints. The outcome of the SDP is a lower bound on the desired value for two reasons: first, because one finds the minimum fidelity within ${\cal Q}_n$, so the fidelity within the quantum set can only be larger; second, because the choice of ${\cal S}$ may not be optimal. For a given choice of ${\cal S}$, one may be able to prove that the SDP bound is tight by exhibiting an explicit quantum strategy which reaches the bound. At the moment of writing, we do not know how to estimate how far from optimal can a choice of ${\cal S}$ be, but the examples shown in this paper demonstrate that intuitive constructions of the swap based on the expected realization of the boxes lead already to much better bounds than the previously reported ones.

The versatility of the method is therefore evident. Having shown that it provides very robust bounds on the most studied example of self-testing, we move to apply it to a case for which no method was previously known: the self-testing of a partially-entangled qutrit state through ternary-outcomes statistics. Later, we shall present also an example of self-testing of measurements; several other examples are presented in \cite{fulllength}.

\textit{Partially-entangled qutrits} - Self-testing of qutrits with ternary measurements, and more generally of box scenarios with more than two outputs per box, was not possible to analyze with Jordan's Lemma \cite{jordan} as used in \cite{miller,reichardt}. With our method, we can achieve it by simply transposing the analysis of the CHSH inequality to the CGLMP inequality $\mathcal{B}_{\mathrm{CGLMP}} \geq 1$~\cite{cglmp}.

The maximum quantum violation of this inequality in the case of three outcomes was conjectured to be $\mathcal{B}_{\mathrm{CGLMP}}(p)=(12-\sqrt{33})/9\approx 0.6950$ \cite{conject}; this was later verified with SDP, up to numerical precision \cite{navascues}. Moreover, it is believed that the maximal quantum violation can only be achieved with the non-maximally entangled state
\begin{align}
	\ket{\overline{\psi}}=\dfrac{1}{\sqrt{2+\gamma^2}} (\ket{00}+\gamma\ket{11}+\ket{22}), \label{cglmpreferencestate}
\end{align}
where $\gamma=(\sqrt{11}-\sqrt{3})/2$. This conjecture will be proved as a corollary of our self-testing.

\begin{figure}[htbp!]
	\begin{center}
	\includegraphics[scale=0.5]{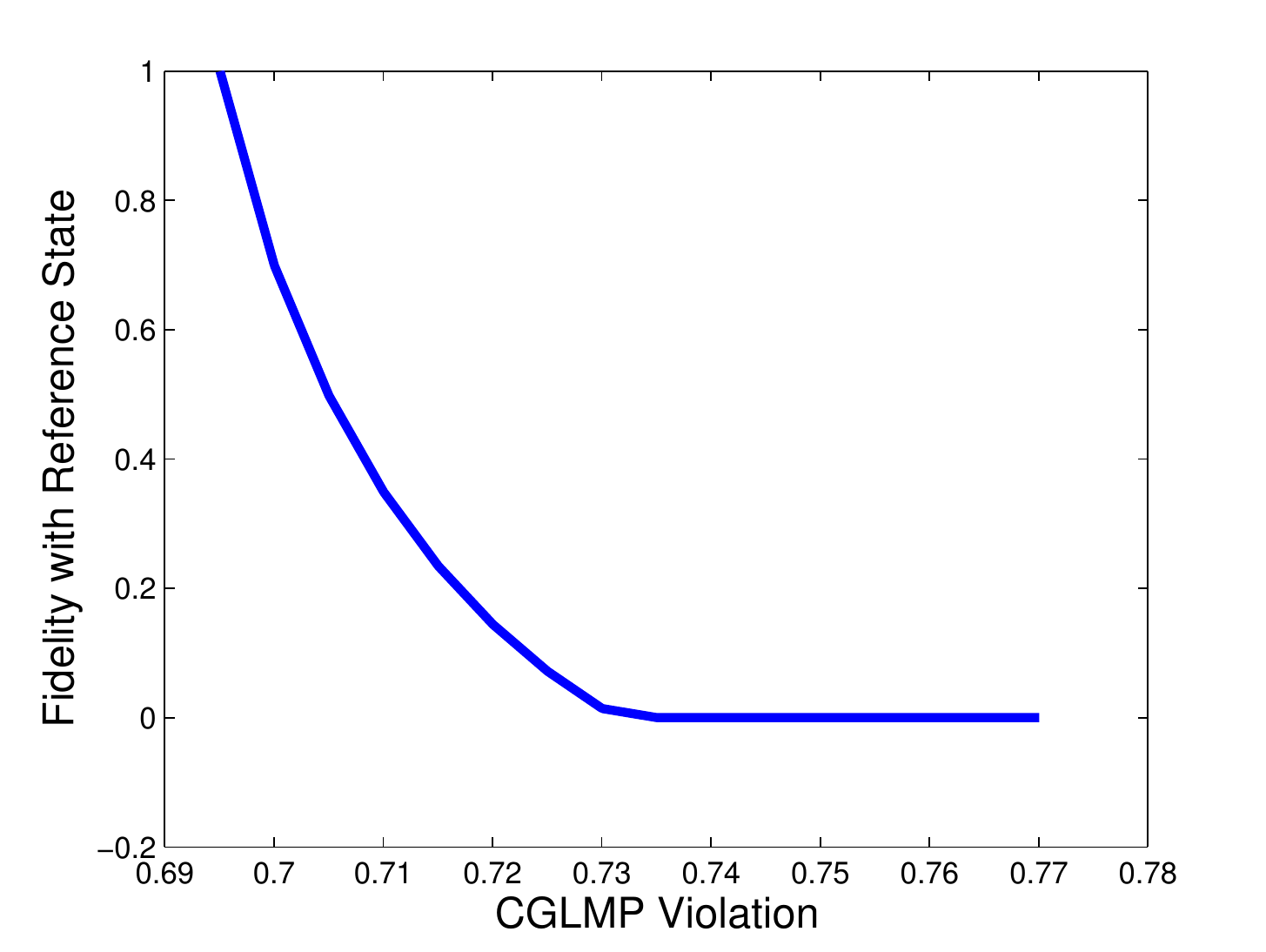}
	\caption{Minimum fidelity of the swapped state with the reference state \eqref{cglmpreferencestate} as a function of the 3-outcome Bell inequality $\mathcal{B}_\mathrm{CGLMP}$.\label{cglmpfidelity}}
\end{center}
\end{figure}

The only technical step consists in finding a suitable ${\cal S}$ for this situation. CNOT operators for qutrit states take a different form than~\eqref{eq:UV}. However, they can still be expressed in terms of the measurement operators $(\overline{E}^x_a, \overline{F}^y_b)$ that yield the maximal CGLMP violation following the technique presented in Appendix A %the Supplemental Material~\cite{suppmat} 
(more details in~\cite{fulllength}). Once this is done, again we obtain the formal expression of the two qutrit swapped state $\rho_\text{swap}$, then we run the SDP to obtain a lower bound on its fidelity with the reference state $\ket{\overline{\psi}}$ as a function of the CGLMP violation. The result is shown in Fig.~\ref{cglmpfidelity}. In particular, the fact that $\bra{\overline\psi}\rho_\text{swap}\ket{\overline\psi}=1$ when the violation is maximal shows that any quantum system violating the CGLMP inequality maximally is indeed unitarily equivalent to $\ket{\overline{\psi}}$.

\textit{Measurement estimation} - As the last application of our method in this paper, we consider certifying measurements rather than states. Suppose that, rather than verifying that $\ket{\psi}$ is close to $\ket{\overline{\psi}}$, we are interested in learning to which degree the actual measurements $\{F^y_b\}$ that Bob's box is performing are well described by some matrices $\{\overline{F}^y_b\}$. The virtual procedure is again based on the intuition of the swap, and thus demonstrates another use of the swap operator $\cal{S}$ introduced earlier: this time consider the task of swapping \textit{into} the box an arbitrary trusted state, then probe the box with different measurements $y$. The figure of merit should quantify how close to the ideal case the boxes perform.

\begin{figure}[htbp!]
	\begin{center}
	\includegraphics[width=0.5\textwidth]{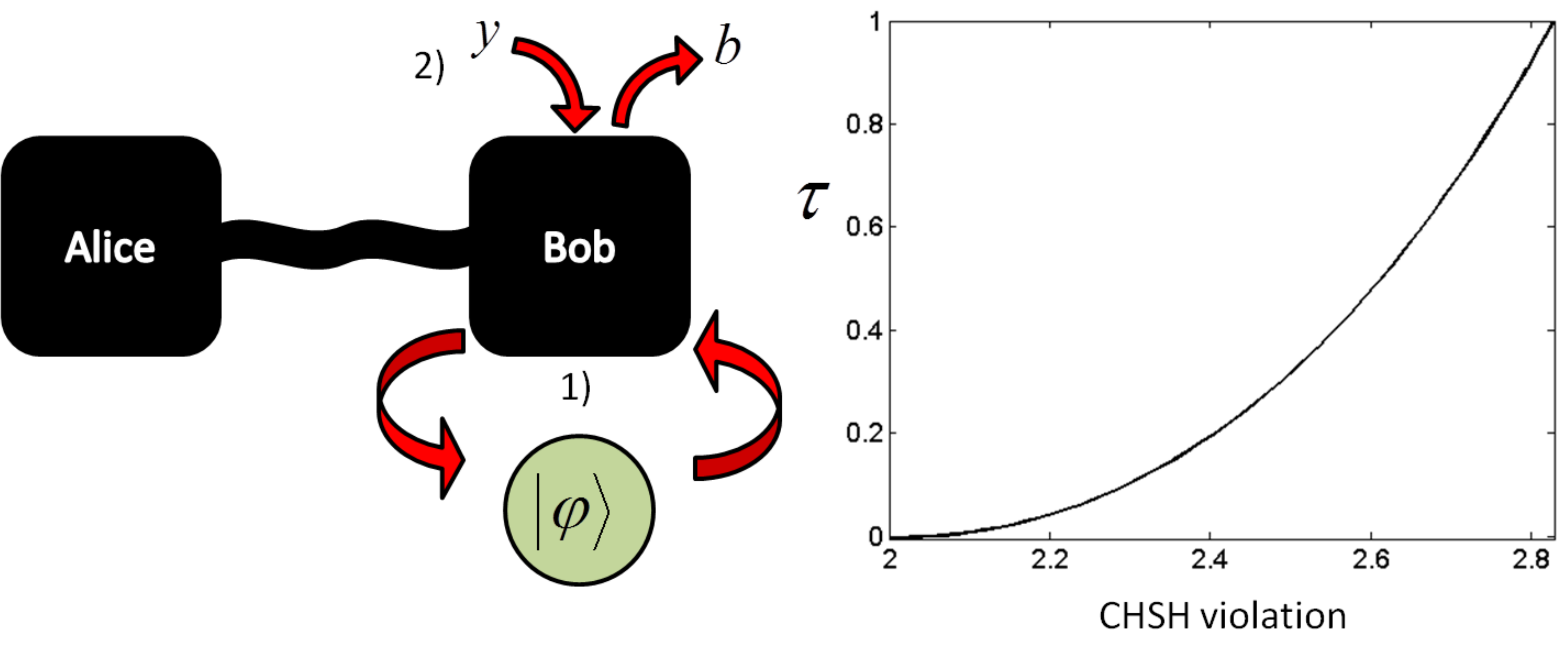}
	\caption{Estimation of Bob's measurements. The protocol works in two steps: 1) We implement a full SWAP of Bob's box and his trusted qubit, that we prepare in state $\ket{\varphi}$. 2) We implement measurement $B_y$ and study the resulting statistics.}
	\label{meas_CHSH}
\end{center}
\end{figure}

For definiteness, let us practice this intuition in the CHSH case (Fig.~\ref{meas_CHSH}, left). We conjecture that Bob's observables are close to $\overline{B}_0=\sigma_z,\overline{B}_1=\sigma_x$. To quantify this hypothesis, we define the figure of merit
\be
\tau\equiv\frac{1}{2}\{P(0|0,0)+P(1|0,1)+P(0|1,+)+P(1|1,-)\}-1,
\ee
where $P(b|y,\varphi)$ denotes the probability of obtaining result $b$ when the trusted qubit was prepared in state $\ket{\varphi}$ and one presses button $y$ after applying the full swap \eqref{eq:swapgen} to Bob's box. $\tau$ is a number ranging from -1 to +1, and $\tau=1$ is achievable only in the ideal case. As before, each $P(b|y,\varphi)$ (and thence $\tau$) is a linear expression in the moments $c$; so a lower bound can be found with the SDP. The result is shown in Fig.~\ref{meas_CHSH}, right, for the case of isotropic boxes. This confirms that Bob's measurement are essentially $\sigma_z$ and $\sigma_x$ when CHSH takes a value close to $2\sqrt{2}$.

\textit{Conclusion} We have described an approach to self-testing that provides much more robust bounds than previously reported and is at the same time very versatile: once the swap operator is constructed, the details of the scenario (ideal cases, figure of merit to be used) enter as parameters. 
The construction of unitaries ${\cal S}$ that provide optimal bounds remains a challenge, but one that can be met with an intuitive understanding of the problem at hand. We have illustrated the power of the method with a few paradigmatic results: the first bound on the singlet fidelity based on CHSH that is robust for real experiments (Fig.~\ref{fig:CHSH}), the first report of self-testing of qutrits using ternary measurements (which also solves a standing conjecture about the kind of states required to violate the CGLMP inequality maximally), and an example of certification of measurements.

\section*{Acknowledgements}
This work is funded by the Singapore Ministry of Education (partly through the Academic Research Fund Tier 3 MOE2012-T3-1-009) and by the National Research Foundation of Singapore. M.N. acknowledges support from the John Templeton Foundation, the European Commission (EC) STREP ”RAQUEL” and the MINECO project FIS2008-01236, with the support of FEDER funds. T.V. acknowledges financial support from a J\'anos Bolyai Grant of the Hungarian Academy of Sciences, the Hungarian National Research Fund OTKA (PD101461), and the T\'AMOP-4.2.2.C-11/1/KONV-2012-0001 project.

%\begin{widetext}

\appendix

\section{The SWAP method for CGLMP}
We present here a more detailed description of the SWAP method for the CGLMP scenario, which was only briefly discussed in the main document. In a forthcoming publication \cite{fulllength}, we will prove that the idea of transferring quantum information from and to the black boxes can be carried even further and generalized to any Bell non-locality scenario with arbitrary number of measurement settings and outcomes.

Here we focus on the CGLMP inequality \cite{cglmp}, which requires two measurement settings on each side, with three possible measurement outcomes. The inequality reads:
\begin{align}
\label{cglmpineq}
	&\mathcal{B}_{\mathrm{CGLMP}}(p)= \nonumber\\
    &p(a<b|x=1,y=1)+p(a>b|x=0,y=1) \nonumber\\
	+&p(a\geq b|x=1,y=0)+p(a<b|x=0,y=0) \ \geq 1.
\end{align}
The maximum quantum violation of the above CGLMP inequality is conjectured \cite{conject} and verified numerically \cite{navascues} to be $\mathcal{B}_{\mathrm{CGLMP}}(p)=(12-\sqrt{33})/9\approx 0.6950$. Moreover, it is believed that the maximal quantum violation can only be achieved with the (non-maximally entangled) state described in \cite{conject,cglmp}. Here we will also prove this conjecture true.

Firstly we give a strategy which is unitarily equivalent to the measurement scheme presented in references~\cite{conject,cglmp} and achieves the maximal violation of CGLMP. The strategy is as follows: Alice's and Bob's first measurements $x,y=0$ correspond to the projectors $\{\proj{0},\proj{1},\proj{2}\}$, namely $\overline{E}_a^0=\proj{a}, \overline{F}_b^0=\proj{b}$. The projectors corresponding to the other measurements $x,y,=1$ are given by $\overline{E}_a^1=\proj{\omega_a}, \overline{F}_b^1=\proj{\omega_b}$, where $\ket{\omega_i}$ and the state to be measured $\ket{\overline{\psi}}$ are as follows
\begin{align}
	\ket{\omega_k} &= \dfrac{1}{3} \left( 2\ket{k} + 2\ket{k+1} -\ket{k+2} \right), \nonumber\\
	\ket{\overline{\psi}} &= \dfrac{1}{3\sqrt{2+\gamma^2}} \left(   (\gamma+\sqrt{3}) (\ket{00}+\ket{11}+\ket{22}) + \right. \nonumber\\
	& \hspace{2.2cm} \gamma (\ket{01}+\ket{12}+\ket{20}) + \nonumber\\
	& \left. \hspace{2.2cm} (\gamma-\sqrt{3}) (\ket{02}+\ket{10}+\ket{21}) \right), \label{cglmpoptimal}
\end{align}

\noindent where all addition above performed inside the kets are modulo 3 and $\gamma = (\sqrt 11 - \sqrt 3)/2$. This strategy up to local unitaries is equivalent to the measurement scheme presented in \cite{conject,cglmp}, which involves complex coefficients.

The above measurements and states of Eq.~(\ref{cglmpoptimal}) shall then be our reference system. Following the method presented in the main document, Alice and Bob will each attach a trusted qutrit initialized in state $\ket{0}$ to the entangled pair in order to certify the state. The next step is to construct the unitary operators which appear in the decomposition of the two-qutrit SWAP operator $S=T U V U$, with $U=\sum_{k=0}^2P^k\otimes\proj{k}$, $V=\sum_{k=0}^2 \proj{k}\otimes P^{-k}$, $T=\id\otimes \sum_{k}\ket{-k}\bra{k}$ and $P=\sum_{k=0}^2\ket{k+1}\bra{k}$. Clearly, we can take $\{E^0_k\}_{k=0}^2$, $\{F^0_k\}_{k=0}^2$ to play the role of the projectors $\{\proj{k}\}_{k=0}^2$ in the first subsystem of the expressions above. A more challenging issue, though, is how to build the translation operator $P$ from the measurement projectors defined in Eq.~(\ref{cglmpoptimal}).

There are many choices to do so; we chose the simplest combination:
\begin{align}
	P &= E^0_0+2E^0_2+\dfrac{1}{2}E^0_1 - \dfrac{3}{2} E^0_0 (2E^1_1+E^1_2) \nonumber\\
	& \hspace{0.2cm} - \dfrac{3}{2}E^0_1 (E^1_1-E^1_2) - \dfrac{3}{2}E^0_2(E^1_1+2E^1_2), \label{cglmpunitary}
\end{align}
which indeed is a translation operator mapping $\ket{0}\rightarrow\ket{1}\rightarrow\ket{2}\rightarrow\ket{0}$ whenever the measurement operators are $E_a^x=\bar{E}_a^x$. Since Alice and Bob's optimal operators are identical, the above formula also applies to Bob's settings if we replace $E$'s by $F$'s.

Note that the choice above in \refereqn{cglmpunitary}, contrary to the CHSH scenario \cite{CHSH}, defines a valid unitary operator only for the optimal strategy of Ref.~\refereqn{cglmpoptimal}. However, in the device independent scenario, when the violation is not optimal, measurement operators can differ from~\refereqn{cglmpoptimal} so that $P$ is not unitary anymore. We address this problem by introducing an extra auxiliary operator, $\hat{P}_A$, which is unitary by construction, and satisfied the constraint that
\begin{equation}\label{eq:defPhat}
\hat{P}_A^\dagger P(E^x_a)\geq 0.
\end{equation}
We then use this operator $\hat{P}$ in the construction of the SWAP instead of $P$, thus ensuring that $S$ is always unitary.

For Bob's side, the swap operators are defined exactly the same way as above for Alice. Thus, we require also another auxiliary operator $\hat{P}_B$. In the SDP, the conditions \eqref{eq:defPhat} for Alice and Bob are relaxed by requiring the positivity of two semidefinite, so-called localizing matrices $\Gamma(\hat{P}_A^\dagger P(E^x_a))$, $\Gamma(\hat{P}_B^\dagger P(F^y_b))$, where $\Gamma$ refers to the moment matrix of \cite{navascues} that proceeds from a quantum realization.

Putting all together, the estimation of the fidelity of the state inside the box $\ket{\psi}$ with respect to the reference state $\ket{\overline{\psi}}$ in Eq.~(\ref{cglmpoptimal}) can be relaxed to the following SDP program:

\begin{align}
	f=\min \;\; &\bracket{\overline{\psi}}{\rho_\text{swap}}{\overline{\psi}} \nonumber\\
	\textrm{such that }&\hspace{0.2cm} c\in {\cal Q}_n \nonumber\\
			    & \sum_{a,b,x,y}{B_{a,b}^{x,y}c_{E_a^xF_b^y}}=\mathcal{B}_\text{CGLMP} \nonumber\\
                & \rho_\text{swap}\ge 0,\quad \tr{\rho_\text{swap}}=1 \nonumber\\
                & \Gamma(\hat{P}_A^\dagger P(E^x_a))\ge 0,\quad \Gamma(\hat{P}_B^\dagger P(F^y_b))\ge 0,
\end{align}
where ${\cal Q}_n$ is a relaxation of the quantum set defined by the positivity of the moment matrix $\Gamma\ge 0$ in a certain level of the NPA hierarchy \cite{navascues}, and $B_{a,b}^{x,y}$ defines the Bell coefficients of the CGLMP inequality in Eq.~(\ref{cglmpineq}).

Notice that here all three semidefinite matrices can be taken real, since, for any feasible point  $\Gamma,\Gamma(\hat{P}_A^\dagger P(E^x_a)),\Gamma(\hat{P}_B^\dagger P(F^x_a))$ of the corresponding complex SDP, the real matrices $\Re\{\Gamma\},\Re\{\Gamma(\hat{P}_A^\dagger P(E^x_a))\},\Re\{\Gamma(\hat{P}_B^\dagger P(F^x_a))\}$ are also positive semidefinite, satisfy the appropriate linear constraints and return the same state fidelity. This is the case because both the figure of merit and the localizing matrices can be expressed as \emph{real} linear combinations of the momenta $c$.

We ran the SDP for various values of $\mathcal{B}_\text{CGLMP}$ for the lowest possible level of the NPA hierarchy which defines all moments appearing in the objective function. The result is shown in Figure~3 of the main document. In particular, the fact that up to numerical precision $\bra{\overline\psi}\rho_\text{swap}\ket{\overline\psi}=1$ when the violation is maximal shows that any quantum system violating the CGLMP inequality maximally is indeed unitarily equivalent to $\ket{\overline{\psi}}$ proving the conjecture of Acin et al.~\cite{conject} true.

%\end{widetext}


\begin{thebibliography}{99}
\bibitem{endnotedet} Here and throughout the whole text, we consider only asymptotic statements, i.e. we assume that the measurement statistics are perfectly reconstructed. Also, it is understood that the observed Bell violation is not fake: the detection loophole must be closed and no-signaling must be assumed or guaranteed.
\bibitem{endnoteuv} Having chosen the ancillas $A'$ and $B'$ in the $\ket{0}$ state, we could have used ${\cal S}_{AA'}=U_{AA'} V_{AA'}$ and the same for Bob, because the first $U$ operator acts trivially. This is effectively equivalent to the isometry used in~\cite{mckague}: therefore, the enormous difference in robustness between that result and ours shows the power of the SDP tool.
\end{thebibliography}
\end{document}